
\input phyzzx

\hyphenation{Schwarz-schild}

\rightline{SU-ITP-93-18}
\rightline{hep-th/9307168}
\rightline{July 1993}

\bigskip\bigskip
\title{String Theory and the Principle of Black Hole Complementarity}

\vfill
\author{Leonard Susskind,\foot{susskind@dormouse.stanford.edu}}
\address{ Department of Physics \break Stanford University \break
            Stanford, CA 94305}
\vfill

\abstract
String theory provides an example of the kind of apparent
inconsistency that
the {\it Principle of Black Hole Complementarity\/} deals with.  To a
freely
infalling observer a string falling through a black hole horizon
appears to be
a Planck size object.  To an outside observer the string and all the
information it carries begin to spread as the string approaches the
horizon.
In a time of order the ``information retention time'' it fills the
entire area
of the horizon.

PACS Categories: 04.60.+n, 12.25.+e, 97.60.Lf, 11.17.+y

\vfill\endpage

\REF\hawk{S.~W.~Hawking,
\journal Comm.~Math.~Phys. & 43 (75) 199
\journal Phys.~Rev. & D14 (76) 2460.}

\REF\done{L.~Susskind, L.~Thorlacius, and J.~Uglum, {\it The
Stretched Horizon
and Black Hole Complementarity}, Stanford University preprint,
SU-ITP-93-15,
hepth@xxx/9305069, June 1993. \hfill\break
\noindent See also G.~`t~Hooft
\journal Nucl. Phys. & B335 (90) 138.}

\REF\dtwo{L.~Susskind and L.~Thorlacius, {\it Gedanken Experiments
Involving
Black Holes}, Stanford University preprint, SU-ITP-93-19, in
preparation.}

\REF\dthree{L.~Susskind
\journal Phys. Rev. & D1 (70) 1182.}

\REF\dfour{M.  Karliner, I. Klebanov and L.  Susskind.
\journal Int. Jour. Mod. Phys. & A3 (88) 1981.}

\noindent I. Introduction

The paradox of information loss in black hole evaporation [\hawk] is
essentially concerned with the localization of information and how it
is
perceived by different observers.  According to the {\it Principle of
Black
Hole Complementarity\/} [\done,\dtwo] no inconsistency follows from
the
following two assumptions:

1)  To a freely falling observer, matter falling toward a black hole
encounters
nothing out of the ordinary upon crossing the horizon.  All quantum
information
contained in the initial matter passes freely to the interior of the
black
hole.

2)  To an observer outside the black hole, matter, upon reaching the
``stretched horizon''[\done], is disrupted and emitted as thermalized
radiation
before crossing the horizon.  All quantum information contained in
the initial
matter is found in the emitted radiation.

In this paper an example will be described in which information
appears to be
localized in extremely different ways to infalling and outside
observers.  The
example relies on the peculiar zero-point fluctuations of fundamental
strings
[\dthree,\dfour].  The result is closely related to the Regge
behavior of
string scattering amplitudes.

Let us first recall a standard argument about why string theory
should not
influence the discussion of black holes and information loss.  It is
widely
understood that strings are extended objects and therefore may
introduce a bit
of nonlocality.  However, the argument goes, the extended objects
have a size
of order the Planck length. On the other hand the horizon of a
massive black
hole is very flat on this scale so that strings can be effectively
replaced by
point particles.  We shall see that this logic is correct for the
freely
falling observer but completely incorrect for external observers.

\noindent II.  The Size of Strings

One of the oldest known and widely ignored properties of strings is
that their
physical size is not well defined unless a ``resolution time'',
$\epsilon,$ is
prescribed [\dthree,\dfour].  The time $\epsilon$ is a smearing time
over which
the internal motions of the string are averaged.  If the resolution
time is
measured in Planck units then the spatial extent of the wave function
of the
string in Planck units satisfies
$$ R^2_{String} \ \sim \ \log {1 \over \epsilon} \;,
\eqn\arestring  $$
for $\epsilon \; << \; 1.$

\noindent Thus as the string is examined with better and better time
resolution
it appears to slowly grow.  For the purposes of low energy physics,
resolution
times are always large and this phenomenon is not important.

Before deriving \arestring \ let us recall that \arestring \ is
closely related
to the well known Regge behavior of string scattering amplitudes.  If
a string
of energy $E \ >> \ 1$ collides with a target at rest, the scattering
amplitude
for momentum transfer $q$ is given by [\dthree]
$$ A(E, \; q^2) \; \sim \ F(q^2)(E)^{-(q^2 \; + \; c)}
\; = \; F(q^2)e^{-(q^2 \; + \; c) \; \ln E}  \;,
\eqn\scattamp  $$
where $c$ is a constant.  Fourier transforming to find the amplitude
as a
function of impact parameter shows that the radius of the scattering
event
grows like \ \ $ (\ln E)^{1 \over 2}$. \ \ If we now assume
(correctly) that
the scattering event averages over a time of order \ \ $\epsilon \; =
\;
E^{-1}$ \ \ we recover \arestring.  The growth of strings with energy
is their
oldest known property.

To derive \arestring \ consider a string in the light cone frame.
The normal
mode expansion for the transverse coordinate of a point $\sigma$ is
$$ X^i(\sigma) \; = \;
X^i_{cm} \; + \; \sum_{l>0}\big(X^i_l\cos{(l\sigma)}
\; + \; \bar X^i_l\sin{(l\sigma)}\big)\; .
\eqn\modexpntion  $$
Consider the mean square transverse distance between the center of
mass and the
material point $\sigma$
$$  \vev{(X(\sigma) \; - \;
X_{cm})^2}\; .  \eqn\transvrse  $$
If the string is in the ground state this reduces to
$$ \vev{(X(\sigma) \; - \;
X_{cm})^2} \; = \;  \sum_l {1 \over l} \; ,
\eqn\reduction  $$
which diverges for every point $\sigma.$  The same divergence is
found in the
mean square distance between any pair of points $\sigma_1,$ and
$\sigma_2.$

If the observation of the string lasts a time $\epsilon$ in the
strings rest
frame the contribution of modes with \ \ $l \; > \; {1 \over
\epsilon}$ \ \ is
averaged out.  The result is
$$ \vev{(X(\sigma) \; - \;
X_{cm})^2}_{\epsilon} \; \equiv \; R^2_{\epsilon} \;
\sim \; \ln {1 \over \epsilon} \; .
\eqn\reslt  $$

Another quantity which diverges as \ $\epsilon \rightarrow 0$ \ is
the total
length of the projection of the string on the transverse plane
[\dfour].  This
is defined by
$$  L \; = \; \int_0^{2 \pi} d \sigma
\Big({\partial X \over \partial \sigma}
{\partial X \over \partial \sigma}\Big)^{1 \over 2}.
\eqn\trnsvrspln  $$
When the resolution time is accounted for one finds that $L$
increases like \
${1 \over \epsilon}.$  \ Because the mean radius $R$ grows so much
slower than
the total length $L$ the string must trace over the same region of
transverse
space many times.  As \ $\epsilon \rightarrow 0$ \ the string becomes
space
filling.  In [\dfour] a particularly graphic illustration of these
facts was
obtained by Monte Carlo sampling of the probability functional of the
string.
We refer the reader to that reference for pictures of typical string
configurations corresponding to decreasing resolution time.

Here we simply remark that as \ $\epsilon \; \rightarrow \; 0$ \ not
only does
the string wave function spread but the information which
distinguishes
different states of the string is also diffused over the area $\sim
\;
R^2_{\epsilon}.$

Now consider a string falling toward a black hole.  An observer
falling with
the string does measurements which we shall suppose involve ordinary
energies
and time scales.  In other words the resolution time in the infalling
frame is
not significantly smaller than the Planck time.  The string and all
its
information is localized in a transverse size of order unity.

Now let us consider an observation of the string done by a distant
fiducial
observer whose clocks register Schwarzchild time.  Suppose the
measurement
again averages over a time \ $\sim \; 1.$ \ But now because of the
red shift
factor, this corresponds to a time in the string frame which is much
smaller.
It is easily seen that the resolution time in the string frame is of
order
$$ \epsilon \; \sim \; \exp -{t \over 4 M}  \; .
\eqn\restime  $$
Accordingly the transverse size of the string seen in such a
measurement is
given by \reslt .  This becomes
$$ R^2 \; \sim \; {t \over M}  \; .
\eqn\nwtransvrs  $$
In other words the distant observer sees the string, upon passing
through the
stretched horizon, start to spread.  In fact the spreading appears to
behave as
if the string was diffusing away from its original transverse
location.

Eventually from the outside point of view the string will fill an
area
comparable to the whole horizon.  This occurs when \ $R^2 \; \sim \;
M^2$ \ or
\ $t \; \sim \; M^3$.

It is interesting that the information retention time defined in
[\dtwo] is
also of order \ $M^3.$ \ This time is defined as follows.  Suppose at
time \ $t
\; = \; 0 $ \ the stretched horizon is in some pure state.  At that
time a
particle in some state \ $\ket{i}$ \ is absorbed at the stretched
horizon.  The
resulting states of the stretched horizon are initially orthogonal
for
different \ $\ket{i}.$ \ However, after some time the density matrix
of the
stretched horizon loses memory of \ $\ket{i}, $ \ the lost
information having
been radiated in the Hawking radiation.  The time for this to occur
is called
the information retention time.  In [\dtwo] it is argued that this
time is of
order \ $M^3.$ \ This suggests the following speculative picture.
The
information in a particle is absorbed at the stretched horizon.
According to
an outside observer it begins to spread as it sinks toward the event
horizon.
At a time  \ $M^3 $ \ it is spread over the entire horizon and can no
longer
expand.  By roughly that time the information must be radiated away.

The same event is viewed by the infalling observer who simply sees a
microscopic string fall past the horizon with nothing to disrupt it
until it
approaches the singularity.

The above description has ignored the splitting and joining of
strings which
can take place near the horizon.  We hope to return to this point at
a later
time.

\noindent III. Acknowledgements

The author thanks L.  Thorlacius for useful discussions and J.
Susskind for
technical assistance.  This work is supported in part by National
Science
Foundation grant PHY89-17438.

\vfill\endpage

\refout
\end